\documentclass[preprint,aps ,nofootinbib]{revtex4}
\usepackage{graphicx}
\usepackage{amsmath}
\usepackage{amsfonts}
\usepackage{amssymb}
\usepackage{color}%
\usepackage{dcolumn}
\providecommand{\U}[1]{\protect\rule{.1in}{.1in}}

\newcommand{\f}{\begin{equation}}
\newcommand{\ff}{\end{equation}}
\newcommand{\fa}{\begin{eqnarray}}
\newcommand{\ffa}{\end{eqnarray}}

\begin{document}
\title{The $3+1$ holographic superconductor with Weyl corrections}
\author{Jian-Pin Wu $^{1}$}
\email{jianpinwu@yahoo.com.cn}
\author{Yue Cao $^{1}$}
\email{yuecao860408@gmail.com}
\author{Xiao-Mei Kuang $^{2}$}
\email{xmeikuang@gmail.com}
\author{Wei-Jia Li $^{1}$}
\email{li831415@163.com}
\affiliation{$^1$Department of Physics, Beijing Normal University, Beijing 100875, China\\$^2$Center for Relativistic
Astrophysics and High Energy Physics, Department of Physics,
Nanchang University, Nanchang 330031, China}

\begin{abstract}

In this paper we study $3+1$ holographic superconductors with Weyl corrections.
We find that the critical temperature of a superconductor with Weyl corrections increases as we amplify the Weyl coupling parameter $\gamma$,
indicating the condensation will be harder when the parameter $\gamma$ decreases.
We also calculate the conductivity and the ratio of gap frequency over critical temperature $\omega_{g}/T_{c}$
numerically for various coupling parameters. We find that the ratio $\omega_g/T_c$
becomes larger when the Weyl coupling parameter $\gamma$ decreases.
We also notice that when $\gamma< 0$ there is an
extra spike that appears inside the gap.

\end{abstract} \maketitle

\section {Introduction}

Since 1986, the discovery that metallic, oxygen-deficient compounds in the Ba-La-Cu-O system can superconduct at $35K$ \cite{HTS},
which is thought as the upper limit allowed by BCS theory \cite{BCS}, has reignited physicists' quest for high-temperature superconductors.
Until now, the high-temperature superconductor materials have been discovered in the laboratory and many common features of the high-temperature superconductors have also been revealed,
but there is no widely accepted theory to explain their properties as the BCS theory in classical superconductors (superconducting below $30K$)\cite{BCS}.
Recently, many theoretical physicists have resorted to the AdS/CFT (Anti-de Sitter/Conformal Field Theory) correspondence \cite{Maldacena1997,Gubser1998,Witten1998,MaldacenaReview}
to offer insight into the pairing mechanism in the high-temperature superconductors.
The AdS/CFT correspondence advances a remarkable equivalence between a $d$-dimensional strongly coupled conformal field theory on the boundary and a $(d+1)$-dimensional weakly coupled dual gravitational description in the bulk.
Through the correspondence, some questions in strongly coupled field theories become amenable to compute and transparent to understand.

The first model for holographic superconductors was established by Hartnoll, Herzog and Horowitz \cite{3H}.
This model originated from the discovery that in an Abelian Higgs model coupled
to gravity in AdS space, Abelian symmetry of Higgs is spontaneously broken due to the existence of a black hole \cite{OriginalDiscovery1,OriginalDiscovery2}.
For a general review of the subject, please see \cite{HorowitzReview,HartnollReview1,HartnollReview2,HerzogReview,McGreevyReview}.
The holographic superconductor is the s-wave superconductors, in which
they considered a charged scalar field and the only Maxwell sector $A=A_{t}$ in an AdS black hole background.
The simple model possesses some basic features
such as the existence of a critical temperature, the occurrence of second order phase transition.

However, there are still many defects in such a simple model.
For instance, at zero temperature and frequencies $\omega\leq \omega_{g}$, the real
part of the conductivity vanishes identically \cite{3H,Hvarious}.
However, as we all know there is a Goldstone boson in our symmetry broken phase,
which may lead to a nonzero conductivity at low frequencies.
As pointed out in Ref.\cite{HHHJHEPSC}, the situation may be caused by the large $N$ limit what we considered in this model.
Therefore, it is interesting to explore the cases beyond the large $N$ limit.
Such investigations are processing, for instance, the holographic superconductors with the higher
curvature corrections \cite{1,2,3,4,5,6,7}.
In addition, many other extending works have been published,
for instance, the holographic superconductor at zero temperature limit \cite{zero1,zero2} or with an external magnetic field \cite{HHHJHEPSC,magnetic1,magnetic2,magnetic3,magnetic4,magnetic5,magnetic6},
the vortex solution \cite{Vortex1,Vortex2,Vortex3,Vortex4},
the holographic superconductor in other gravity theories \cite{otherG1,otherG2,otherG3,otherG4,otherG5,otherG6,otherG7,otherG8},
a general class of holographic superconductors \cite{stuckelberg1,stuckelberg2,stuckelberg3,stuckelberg4,stuckelberg5} and a holographic superconductor model in the Born-Infeld electrodynamics \cite{BI}.
Later both p-wave and d-wave holographic superconductors were also constructed \cite{pd1,pd2,pd3,pd4,pd5}.

In order to explore the effects beyond the large $N$ limit on the holographic superconductor,
we will take into account an interesting extension of Einstein-Maxwell theory in this paper,
which involves a coupling between the Maxwell field and the bulk Weyl tensor \cite{WeylM}.
In Ref. \cite{WeylM}, the authors calculated the conductivity and charge diffusion with Weyl corrections and
showed that corrections break the universal relation with the $U(1)$ central charge observed at leading order.
It is interesting to see when including Weyl corrections, how the picture of the holographic superconductor is modified.

Our paper is organized as follows. In section II,
we construct the $3+1$ holographic superconductor with Weyl corrections and present numerical results
for the condensation and critical temperature of the holographic superconductor.
Then we investigate the conductivity numerically in section III.
Conclusions and discussions follow in section IV.
Finally, a derivation of the retarded Green's function is presented in appendix \ref{appBoundaryConditions}.

\section {The $3+1$ holographic superconductor with Weyl corrections}

The simplest $3+1$ holographic superconductor model includes the gravitational field, the $U(1)$ gauge field, and the complex
scalar field of mass $m$ and charge $q$. The corresponding action is given by \cite{3H,Hvarious}
\begin{eqnarray}
\label{action}
S=\int d^{5}x \sqrt{-g}\left\{\frac{1}{16\pi G_{N}}\left(R+\frac{12}{L^{2}}\right)
+\frac{1}{q^{2}}\mathcal{L}_{m}\right\},
\end{eqnarray}
with
\begin{eqnarray}
\label{matterL}
\mathcal{L}_{m}=-\left(\frac{1}{4} F^{\mu\nu} F_{\mu\nu}
+\frac{1}{L^{2}} |D_{\mu} \Psi|^{2} + \frac{m^{2}}{L^{4}} |\Psi|^{2}\right).
\end{eqnarray}
where $G_{N}$ is the gravitational Newton constant,
the $12/L^{2}$ term gives a negative cosmological constant and $L$ is the AdS radius.
Moreover we have introduced $F_{\mu\nu}=\partial_{\mu}A_{\nu}-\partial_{\nu}A_{\mu}$
and $D_{\mu}=\nabla_{\mu}-i A_{\mu}$.

In this paper, we aim at a holographic superconductor with Weyl corrections \cite{WeylM}.
In this case, the matter Lagrangian density $\mathcal{L}_{m}$
is replaced by \cite{WeylM}
\begin{eqnarray}
\label{matterLWeyl}
\mathcal{L}_{mW}=-\left[\frac{1}{4}\left(F^{\mu\nu} F_{\mu\nu}-4\gamma C^{\mu\nu\rho\sigma}F_{\mu\nu}F_{\rho\sigma}\right)
+\frac{1}{L^{2}} |D_{\mu} \Psi|^{2} + \frac{m^{2}}{L^{4}} |\Psi|^{2}\right].
\end{eqnarray}
where $\gamma$ is a (dimensionful) constant and $C_{\mu\nu\rho\sigma}$ is the Weyl tensor.
There is a limit on the parameter $\gamma$ \cite{WeylM}:
\begin{eqnarray}
\label{gamma}
-\frac{L^{2}}{16}<\gamma<\frac{L^{2}}{24},
\end{eqnarray}
the upper bound is because there exists an additional singular
point when $\gamma=\frac{L^{2}}{24}$ and the lower bound is due to the causality constraints.

We will work in the probe approximation ($G_{N}\rightarrow 0$ and $g^{2}\rightarrow 0$) \cite{3H,Hvarious},
where the gravity sector is effectively decoupled from the matter sector. Without loss of generality, we fix $q = 1$.
In this limit, the metric is given by an AdS-Schwarzschild black hole:
\begin{eqnarray}
\label{metric1}
ds_{5}^{2}=\frac{r^{2}}{L^{2}}(-f(r)dt^{2}+dx_{i}dx^{i})+\frac{L^{2}}{r^{2}f(r)}dr^{2},
\end{eqnarray}
where
\begin{eqnarray}
\label{f1}
f(r)=1-\left(\frac{r_{H}}{r}\right)^{4},
\end{eqnarray}
with $r_{H}>0$ the horizon and $r\rightarrow \infty$ boundary of the bulk.
The Hawking temperature of the black hole is determined by the Schwarzschild radius:
\begin{eqnarray}
\label{HawkingT}
T=\frac{r_{H}}{\pi L^{2}},
\end{eqnarray}
which is also the temperature of the conformal field theory on the boundary of the AdS spacetime.

%
%

Applying the Euler-Lagrange equation, we can derive the equations of motion for the scalar field
\begin{eqnarray}
\label{ScalarEOM}
D_{\mu}D^{\mu}\Psi-m^{2}\Psi=0,
\end{eqnarray}
and the generalized Maxwell＊s equations
\begin{eqnarray}
\label{GMaxwellEOM}
\nabla_{\mu}(F^{\mu\nu}-4\gamma C^{\mu\nu\rho\sigma} F_{\rho\sigma})=i[\Psi^{\ast}D^{\nu}\Psi-\Psi D^{\nu\ast}\Psi^{\ast}],
\end{eqnarray}
where $C_{\mu\nu\rho\sigma}$ is the Weyl tensor and has
the following nonzero components in $AdS_{5}$,
\begin{eqnarray}
\label{WeylT}
C_{0i0j}=\frac{f(r)r_{H}^{4}}{L^{6} } \delta_{ij},~~
C_{0r0r}=-\frac{3 r_{H}^{4}}{L^{2} r^{4}},~~
C_{irjr}=-\frac{r_{H}^{4}}{L^{2} r^{4} f(r)} \delta_{ij},~~
C_{ijkl}=\frac{r_{H}^{4}}{L^{6}} \delta_{ik} \delta_{jl}.
\end{eqnarray}

For convenience, in the following analysis, we will set $L=1$. Consider the following Ansatz:
\begin{eqnarray}
\label{ansatz}
\Psi=\Psi(r),~~~ A_{t}=\Phi(r),
\end{eqnarray}
then the equations of motion (\ref{ScalarEOM}) and (\ref{GMaxwellEOM}) reduce to
\begin{eqnarray}
\label{ScalarEOM1}
\Psi''+\left(\frac{f'}{f}+\frac{5}{r}\right)\Psi'+\frac{\Phi^{2}\Psi}{r^{4} f^{2}} -\frac{m^{2}\Psi}{r^{2}f}=0
\end{eqnarray}
\begin{eqnarray}
\label{GMaxwellEOM1}
\left(1-\frac{24\gamma r_{H}^{4}}{r^{4}}\right)\Phi''
+\left(\frac{3}{r}+\frac{24\gamma r_{H}^{4}}{r^{5}}\right)\Phi'
-\frac{2\Phi \Psi^{2}}{r^{2} f}
=0,
\end{eqnarray}
where the prime represents derivative with respect to $r$.
It is more convenient to work in terms of the coordinate $z=r_{H}/r$,
in which $z=1$ is the horizon and $z=0$ the boundary at infinity.
Then the the equations of motion (\ref{ScalarEOM1}) and (\ref{GMaxwellEOM1})
can be reexpressed as
\begin{eqnarray}
\label{ScalarEOM2}
\Psi''
-\frac{3+z^{4}}{z(1-z^{4})}\Psi'
+\frac{1}{r_{H}^{2} (1-z^{4})^{2}} \Phi^{2}\Psi
-\frac{m^{2}}{z^{2}(1-z^{4})}\Psi=0
\end{eqnarray}
\begin{eqnarray}
\label{GMaxwellEOM2}
(1-24\gamma z^{4})\Phi''
-\left(\frac{1}{z}+72\gamma z^{3}\right)\Phi'
-\frac{2}{z^{2}(1-z^{4})}\Psi^{2}\Phi
=0,
\end{eqnarray}
where prime now denotes derivative with respect to $z$.
We next examine the asymptotic behavior of the fields close to the AdS boundary.
We first fix the mass of the scalar field to $m^{2}=-3$, which is above the Breitenlohner-Freedman
bound\cite{BF}, so that $\Psi$ will have asymptotic behavior
\begin{eqnarray}
\label{BconditionI1}
\Psi=\Psi_{-}z+\Psi_{+}z^{3}.
\end{eqnarray}
For $m^{2}\geq -d^{2}/4+1$, the scalar
condensation is given by $\langle\mathcal{O}_{+}\rangle=\Psi_{+}$.
$\Psi_{-}$ is dual to a source for the operator.
In order to have spontaneous symmetry breaking, we choose the operator to condense without
being sourced.
On the other hand, the asymptotic expansion of $\Phi$ has the following form
\begin{eqnarray}
\label{BconditionI2}
\Phi=\mu-\rho z^{2},
\end{eqnarray}
where $\mu$ and $\rho$ are the chemical potential and charge density, respectively.
Furthermore, at the horizon $z=1$, the regularity gives two conditions:
\begin{eqnarray}
\label{BconditionH1}
\Psi'(1)=\frac{2}{3}\Psi(1)
\end{eqnarray}
\begin{eqnarray}
\label{BconditionH2}
\Phi(1)=0.
\end{eqnarray}

\begin{figure}
\center{
\includegraphics[scale=1]{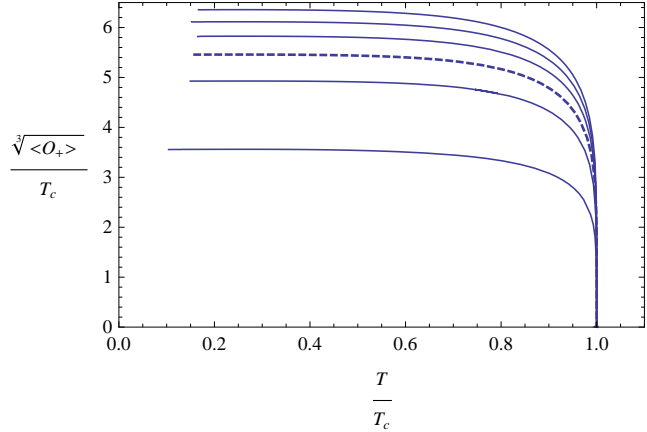}\hspace{1cm}
\caption{\label{Condense1} The condensation as a function of temperature for the operators $<\mathcal{O}_{+}>$.
$\gamma=-0.06, -0.04, -0.02, 0, 0.02, 0.04$ from top to bottom and the dotted line is just the case $\gamma=0$.}}
\end{figure}

Combining with the boundary conditions (\ref{BconditionI1}), (\ref{BconditionI2}), (\ref{BconditionH1}) and (\ref{BconditionH2}), we solve equations of motion (\ref{ScalarEOM2}) and (\ref{GMaxwellEOM2}) numerically by using a shooting method
and plot FIG.\ref{Condense1} to demonstrate
the condensation as a function of temperature for the operators $<\mathcal{O}_{+}>$.
The curve in FIG.\ref{Condense1} is qualitatively similar to that obtained in the holographic superconductors in the Einstein-Maxwell electrodynamics \cite{Hvarious}, where the condensation $<\mathcal{O}_{+}>$ goes to a constant at zero temperature.

Furthermore, from FIG.\ref{Condense1}, it is easy to find that when including the Weyl corrections, this model shows that the critical
temperatures will be higher and higher as the parameter $\gamma$ changes from $-0.06$ to $0.04$.
This means that, in the holographic superconductor with Weyl corrections,
the critical temperature becomes smaller and
the scalar hair is formed harder when $\gamma< 0$, but the cases are just opposite when $\gamma> 0$.

In table (\ref{Tc}) we present the critical temperature $T_{c}$ for the condensations with different values of parameter $\gamma$.
From the table, we can infer that if we relax the causality condition imposed on the low bound of $\gamma$,
the critical temperature will become small and eventually vanish by minimizing the coupling parameter $\gamma$.
Therefore, when the Weyl coupling parameter $\gamma$ is enough small, the Weyl corrections will spoil the holographic superconductor.

\begin{widetext}
\begin{table}[ht]
\begin{center}

\begin{tabular}{|c|c|c|c|c|c|c|}
         \hline
$~~\gamma~~$ &~~$-0.06$~~&~~$-0.04$~~&~~$-0.02$~~&~~$0$~~&~~$0.02$~~&~~$0.04$~~
          \\
        \hline
~~$T_{c}$~~ & ~~$0.170\rho^{1/3}$~~ & ~~$0.177\rho^{1/3}$~~ & ~~$0.185\rho^{1/3}$~~& ~~$0.198\rho^{1/3}$~~& ~~$0.219\rho^{1/3}$~~& ~~$0.304\rho^{1/3}$~~
          \\
        \hline
\end{tabular}
\caption{\label{Tc} The critical temperature $T_{c}$ for different values of Weyl coupling parameter $\gamma$.}
\end{center}
\end{table}
\end{widetext}

\section {Electrical conductivity}

In this section, we will calculate the electrical conductivity $\sigma$ as a function of frequency $\omega$.
According to the AdS/CFT dictionary, the conductivity is obtained by solving for the perturbation
of vector potential $A_{x}$ in the bulk. Without loss of generality,
we will consider just the conductivity in the $x$-direction.
Furthermore, for simplicity we will work at zero momentum ($k=0$) and suppose that the perturbation of vector potential
is radially symmetric and has a time dependence as $\delta A_{x}(t,r)=A_{x}(r)e^{-i\omega t}dx$.
Applying the generalized Maxwell equations (\ref{GMaxwellEOM}),
we have the following equation of motion for the perturbation of vector potential $\delta A_{x}(t,r)$:
\begin{eqnarray}
\label{GMaxwellxEOM}
(1&+&\frac{8\gamma r_{H}^{4}}{r^{4}})A''_{x}
+\left[\left(\frac{f'}{f}+\frac{3}{r}\right)+ \frac{8\gamma r_{H}^{4}}{r^{4}}\left(\frac{f'}{f}-\frac{1}{r}\right)\right]A'_{x}
\nonumber\\
&& \quad
+\left(1+\frac{8\gamma r_{H}^{4}}{r^{4}}\right)\frac{\omega^{2}}{r^{4}f^{2}}A_{x}
-\frac{2A_{x}\Psi^{2}}{r^{2} f}
=0.
\end{eqnarray}

The causal propagation on the boundary requires the ingoing wave boundary conditions at the horizon \cite{Son}
\begin{eqnarray}
\label{ingoingC}
A_{x}(r)\sim (r^{2}f(r))^{-i\frac{\omega}{4r_{H}}},
\end{eqnarray}
this condition gives rise to the retarded Green function.
By the large $r$ expansion of Eq.(\ref{GMaxwellxEOM}), we can obtain the general solution at large radius ($r\rightarrow \infty$) in $AdS_{5}$,
\begin{eqnarray}
\label{Axinfty}
A_{x}(r)=A^{(0)}+\frac{A^{(2)}}{r^{2}}+\frac{A^{(0)}\omega^{2}}{2}\frac{\log \Lambda r}{r^{2}}+\ldots
\end{eqnarray}
where $A^{(0)}$, $A^{(2)}$ and $\Lambda$ are integration constants.

From linear response theory, we know that the conductivity $\sigma$ is related to the retarded current-current two-point function for
global $U(1)$ symmetry,
\begin{eqnarray}
\label{conductivityD}
\sigma(\omega)=\frac{1}{i\omega}G^{R}(\omega,k=0).
\end{eqnarray}

Using the standard AdS/CFT technique, the retarded current Green's function $G^{R}$ can be calculated as\footnote{In Appendix \ref{appBoundaryConditions},
we will give a detailed derivation of the retarded Green's function, following Ref.\cite{Son}. A similar calculation can be also found in Ref.\cite{Myers}.
One find that Weyl term has no explicit effect on the retarded Green's function and
so we get the same expression of the retarded Green's function in Einstein theory for a standard Maxwell field.
Therefore, in the following calculation, we can safely use the Eq.(\ref{conductivityD}).}
\begin{eqnarray}
\label{conductivityD}
G^{R}(\omega)=-\lim_{r \rightarrow \infty} r^{3}f(r)\frac{\partial_{r}A_{x}(r,\omega)}{A_{x}(r,\omega)},
\end{eqnarray}
where $A_{x}$ is purely in-falling at the horizon.
Substituting the solution (\ref{Axinfty}) into the above equation,
one can easily obtain the explicit expression of the retarded Green's function
\begin{eqnarray}
\label{GreenFAdS5}
G^{R}(\omega)=2\frac{A^{(2)}}{A^{(0)}}+\omega^{2}(\log \Lambda r - \frac{1}{2}).
\end{eqnarray}

The logarithmic term suffers from an ambiguity scale $\Lambda$,
which leads to a divergence in the Green's function that has to be removed by an appropriate counterterm,
as explained in \cite{Hvarious}.
After adding a counterterm to cancel the $\log \Lambda r$, we can reexpressed the retarded Green's function as following
\begin{eqnarray}
\label{GreenFAdS5v1}
G^{R}(\omega)=2\frac{A^{(2)}}{A^{(0)}}-\frac{\omega^{2}}{2}.
\end{eqnarray}

Therefore, the conductivity is given by
\begin{eqnarray}
\label{conductivityE}
\sigma(\omega)=-\frac{i A^{(2)}}{\omega A^{(0)}}+\frac{i\omega}{2}.
\end{eqnarray}

In order to obtain the $A^{(0)}$ and $A^{(2)}$, we must solve Eq.(\ref{GMaxwellxEOM}) numerically with the boundary condition(\ref{ingoingC}).
It is also convenient to work in the coordinate $z=r/r_{H}$, in which Eq. (\ref{GMaxwellxEOM}) becomes
\begin{eqnarray}
\label{GMaxwellxEOM1}
(1&+&8\gamma z^{4})A''_{x}
-\frac{1+z^{4}(3-24\gamma)+56\gamma z^{8}}{z(1-z^{4})}A'_{x}
\nonumber\\
&& \quad
+\frac{\left(1+8\gamma z^{4}\right)\omega^{2}}{(1-z^{4})^{2}}A_{x}
-\frac{2r_{H}^{2}A_{x}\Psi^{2}}{z^{2}(1-z^{4})}
=0.
\end{eqnarray}

\begin{figure}
\center{
\includegraphics[scale=0.9]{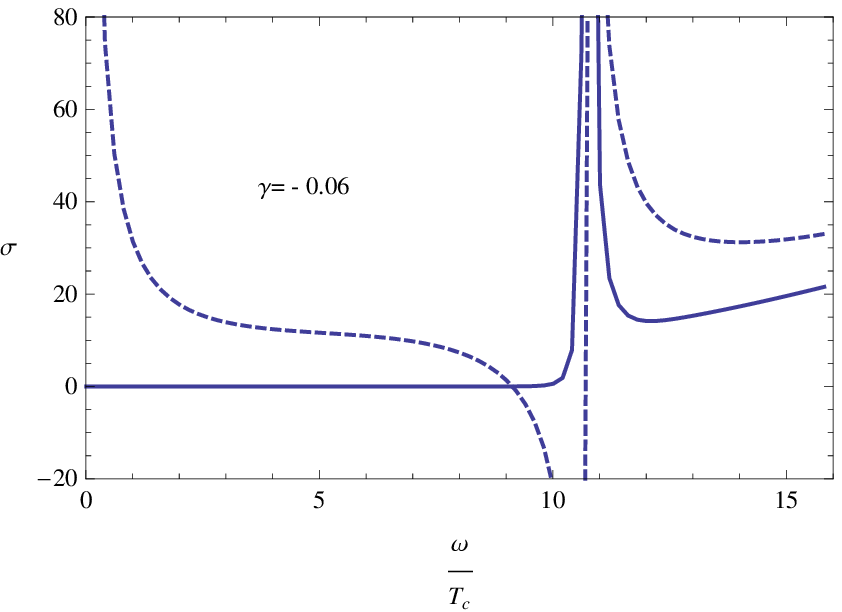}\hspace{1cm}
\includegraphics[scale=0.9]{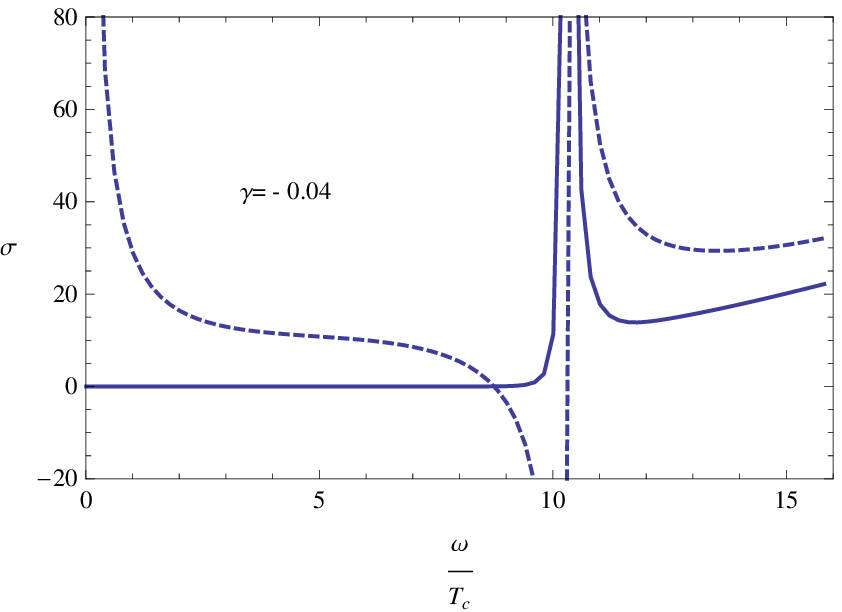}\\ \hspace{1cm}}
\center{
\includegraphics[scale=0.9]{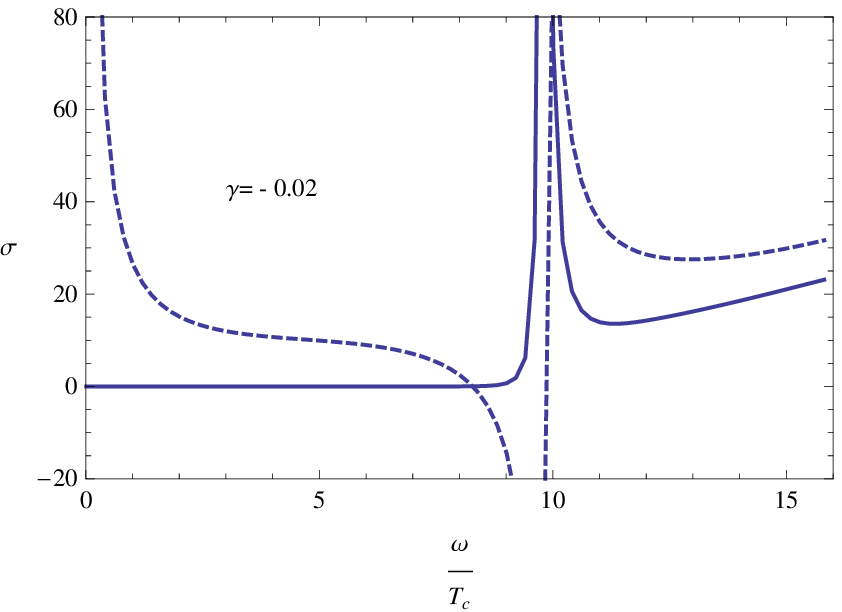}\hspace{1cm}
\includegraphics[scale=0.9]{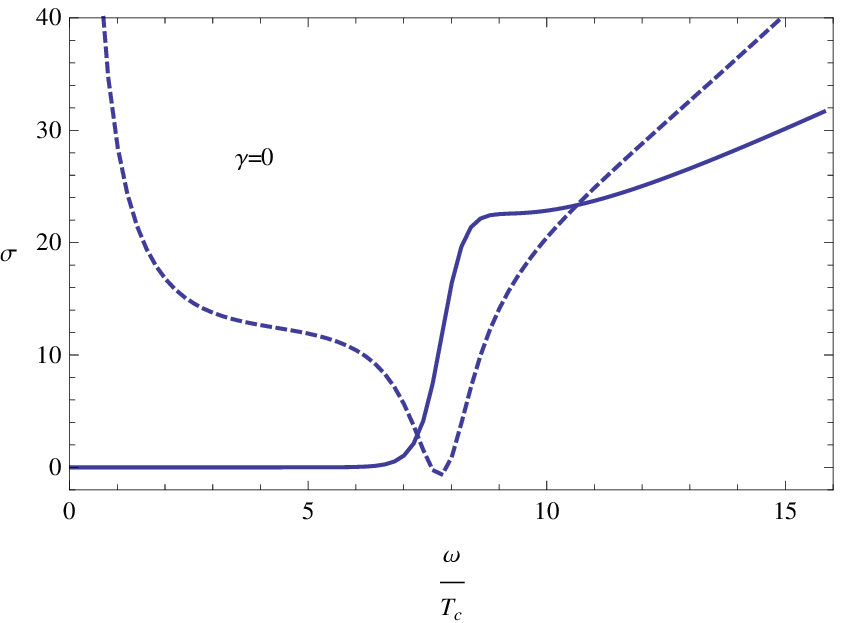}\\ \hspace{1cm}}
\center{
\includegraphics[scale=0.9]{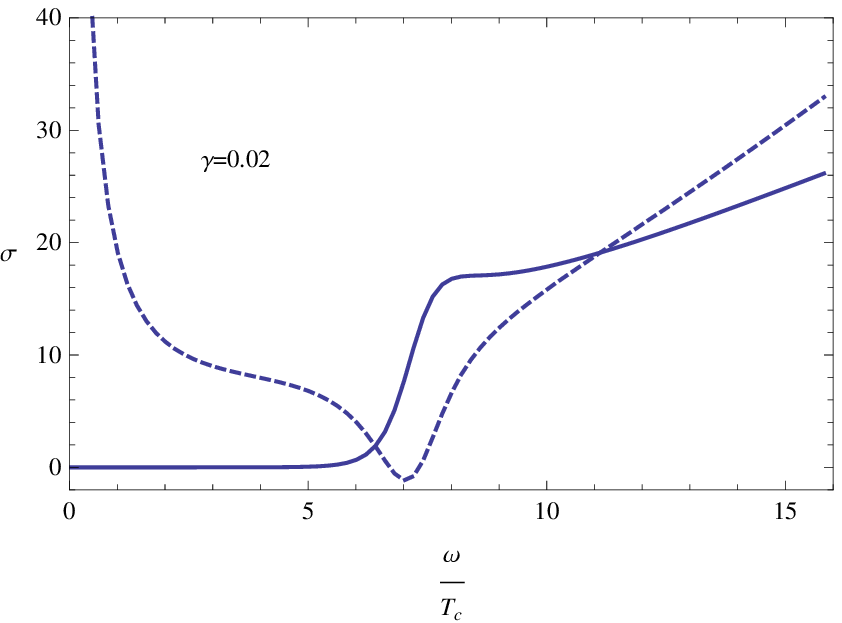}\hspace{1cm}
\includegraphics[scale=0.9]{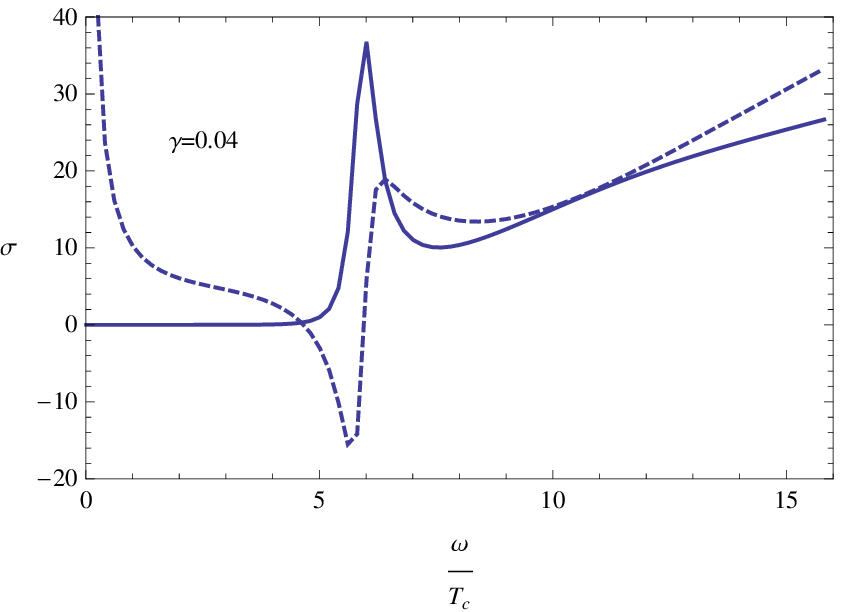}\\ \hspace{1cm}
\caption{\label{conductivity} Conductivity for superconductors with different values of $\gamma$ at the same
temperature $T/T_{c}\approx 0.3$. The solid line represents the real part, and dotted line the imaginary part. There is a
extra spike that appears inside the gap when $\gamma< 0$.}}
\end{figure}

The numerical results of the frequency dependent conductivity with different values of $\gamma$ at the same
temperature $T/T_{c}\approx 0.3$ are illustrated in FIG.\ref{conductivity},
where the solid line represents the real part, and dotted line the imaginary part of $\sigma$.

From FIG.\ref{conductivity}, we find that the conductivity for superconductors with Weyl corrections shares two main common features with the standard version\cite{3H,Hvarious}:

(1)The imaginary part has a pole at $\omega=0$, which indicates that the real part contains a delta function according to the Kramers-Kronig relation.

(2)There exists a gap in the conductivity. The conductivity rises quickly near the gap frequency $\omega_{g}$.

However, after introducing the Weyl corrections, we find that there is an interesting phenomenon that the
ratio of gap frequency over critical temperature $\omega_{g}/T_{c}$ is unstable and
running with the parameter $\gamma$ as the cases in Gauss-Bonnet gravity \cite{1,2} or Quasi-topological Gravity \cite{4}.

Furthermore, we find that the ratio $\omega_{g}/T_{c}$ increases with the fall of the coupling parameter $\gamma$.
In addition, the ratio $\omega_{g}/T_{c} > 8$ when $\gamma < 0$, and the cases are just opposite when $\gamma > 0$.
This is very different from the cases in Gauss-Bonnet gravity \cite{1,2} or Quasi-topological Gravity \cite{4},
in which $\omega_{g}/T_{c}$ is always greater than $8$.

Finally, we also notice that when $\gamma< 0$ there is an
extra spike that appears inside the gap.
a spike has been also observed in the usual holographic superconductor when $m^{2}$ approaches to  the BF bound \cite{Hvarious,HorowitzReview}.

\section{Conclusions and discussion}

In the above sections, we have constructed the holographic superconductor with the Weyl corrections and obtained the numerical solutions of this model.
In the parameter space of the Weyl coupling $-\frac{L^{2}}{16}<\gamma<\frac{L^{2}}{24}$, we find that the critical
temperature will be higher as we amplify the parameter $\gamma$.
Therefore, the condensation gets harder when $\gamma< 0$, which is
consistent with the fact that the higher derivative corrections suppress
the phase transition as obtained in Ref. \cite{1,2,4}.
While $\gamma> 0$, the result is just opposite.
In addition, if we relax the causality condition imposed on $\gamma$,
the Weyl corrections will spoil the holographic superconductor when the Weyl coupling parameter $\gamma$ is enough small.

Finally, we calculate the conductivity of holographic superconductors numerically and find
the ratio $\omega_g/T_c$ is
unstable and becomes larger when the Weyl coupling parameter $\gamma$ decreases.
We also notice that when $\gamma< 0$ there is an
extra spike that appears inside the gap, which has been also observed in the usual holographic superconductor when $m^{2}$ approaches the BF bound.

This work is the first step towards studying the Weyl correction effects on the holographic superconductor.
Obviously, the next step is to consider the  backreaction, which may be particularly important when the coupling parameter is positive and large.
Another important work is to test Coleman-Mermin-Wagner theorem in holographic superconductor \cite{CMW}.
Therefore, we will go on exploring this case in the future works.

\begin{acknowledgments}

We are grateful to Sean A. Hartnoll, Christopher P. Herzog, Sugumi Kanno, Adam Ritz, Yi Ling, and Hongbao Zhang for reply and useful discussions.
J. P. Wu and Y. Cao are partly supported by NSFC(No.10975017).
X. M. Kuang is partly supported by NSFC(Nos.10663001, 10875057), JiangXi SF(Nos.0612036, 0612038), Fok Ying Tung
Education Foundation(No.111008) and the key project of Chinese Ministry of Education(No.208072).
He also acknowledges the support by the Program for Innovative Research Team
of Nanchang University.
W. J. Li is partly supported by NSFC (No.10975016).

\end{acknowledgments}

\begin{appendix}

\section{The Retarded Green's function}\label{appBoundaryConditions}

In this appendix, following the method of Ref.\cite{Son}, we calculate the retarded current Green's function $G^{R}$ by using the AdS/CFT correspondence.
We focus on the gauge field contribution.
Evaluating the action of gauge field with the Weyl correction on-shell:
\begin{eqnarray}\label{a}
S&=&\int d^5x\sqrt{-g}[-\frac{1}{4}(F_{\mu\nu}F^{\mu\nu}-4\gamma C^{\mu\nu\rho\sigma}F_{\mu\nu}F_{\rho\sigma})]
\nonumber\\
&=&\int d^5x\sqrt{-g}[\frac{1}{4}\nabla_\mu F^{\mu\nu}A_\nu-\frac{1}{4}\nabla_\mu
(F^{\mu\nu}A_\nu)-\frac{1}{4}\nabla_\nu F^{\mu\nu}A_\mu+\frac{1}{4}\nabla_\nu(F^{\mu\nu}A_\mu)\nonumber\\&+&
\gamma\nabla_\mu(C^{\mu\nu\rho\sigma}A_\nu F_{\rho\sigma})-\gamma\nabla_\mu(C^{\mu\nu\rho\sigma}F_{\rho\sigma})A_\nu-\gamma\nabla_\nu(C^{\mu\nu\rho\sigma}A_\mu F_{\rho\sigma})+\gamma\nabla_\nu(C^{\mu\nu\rho\sigma}F_{\rho\sigma})A_\mu]
\nonumber\\
&=&-\int d^5x\sqrt{-g}[\frac{1}{2}\nabla_\mu
(F^{\mu\nu}A_\nu)-2\gamma\nabla_\mu(C^{\mu\nu\rho\sigma}F_{\rho\sigma}A_\nu)]
\nonumber\\
&+&\int d^5x\sqrt{-g}[\frac{1}{2}\nabla_\mu
(F^{\mu\nu})A_\nu-2\gamma\nabla_\mu(C^{\mu\nu\rho\sigma}F_{\rho\sigma})A_\nu]\nonumber\\
&=&-\int d^5x\sqrt{-g}[\frac{1}{2}\nabla_\mu
(F^{\mu\nu}A_\nu)-2\gamma\nabla_\mu(C^{\mu\nu\rho\sigma}F_{\rho\sigma}A_\nu)]\nonumber\\
&=&-\int_{\partial\mathcal {M}} d^4x\sqrt{-h}[\frac{1}{2}F^{\mu\nu}n_\mu A_\nu-2\gamma C^{\mu\nu\rho\sigma}F_{\rho\sigma}n_\mu A_\nu],
\end{eqnarray}
we find that the action reduces to a surface term due to the bulk contribution vanishing.
Furthermore, the above action can be explicitly written as
\begin{eqnarray}\label{b}
S&=&-\frac{1}{2}\int_{\partial\mathcal {M}} d^4x\sqrt{-h}g^{rr}g^{xx}\left(1-8\gamma g^{rr}g^{xx} C_{rxrx}\right)n_{r}A_{x}\partial_{r}A_{x}
\nonumber\\
&=&-\frac{1}{2}\int_{\partial\mathcal {M}} d^4x r^{3} f(r)\left(1+\frac{8\gamma}{r^{4}}\right)A_{x}\partial_{r}A_{x}
\end{eqnarray}

Near the boundary ($r\rightarrow \infty$), the Weyl term can be neglected and the action reduces to the usual case
\begin{eqnarray}\label{c}
S=-\frac{1}{2}\int_{\partial\mathcal {M}} d^4x r^{3} f(r)A_{x}\partial_{r}A_{x}\mid_{r\rightarrow \infty}
\end{eqnarray}

Transforming Eq.(\ref{c}) to the Fourier space and comparing the result with the standard AdS/CFT case
\begin{eqnarray}\label{d}
S=\frac{1}{2}\int_{\partial\mathcal {M}} \frac{d^4\mathbf{k}}{(2\pi)^{4}} A_{x}(-\mathbf{k})G^{R}(\mathbf{k})A_{x}(\mathbf{k})\mid_{r\rightarrow \infty}
\end{eqnarray}
we can get the expression of the retarded Green's function
\begin{eqnarray}\label{e}
G^{R}(\mathbf{k})=-\lim_{r \rightarrow \infty} r^{3}f(r)\frac{A_{x}(r,-\mathbf{k})\partial_{r}A_{x}(r,\mathbf{k})}{A_{x}(r,-\mathbf{k}) A_{x}(r,\mathbf{k})}
\end{eqnarray}
where $\mathbf{k}=(\omega,\vec{k})$. Working at zero momentum ($\vec{k}=0$), the retarded Green's function can be furthermore expressed as
\begin{eqnarray}\label{f}
G^{R}(\omega)=-\lim_{r \rightarrow \infty} r^{3}f(r)\frac{\partial_{r}A_{x}(r,\omega)}{A_{x}(r,\omega)}
\end{eqnarray}

The expression of the retarded Green's function is the same as that of Einstein theory for a standard Maxwell field,
$i.e.$, Weyl term has no effect on the retarded Green's function.
If $A_{x}$ is normalized to $A_{x}(r\rightarrow \infty)=1$, the retarded Green's function is often expressed as\cite{Hvarious}
\begin{eqnarray}
\label{g}
G^R = - \lim_{r \rightarrow \infty} r^{3}f(r) A_x A_x'.
\end{eqnarray}

\end{appendix}

\end{document}